\documentclass[
  aps,
  prd,
  onecolumn,
  superscriptaddress,
  nofootinbib
]{revtex4-2}

\pdfoutput=1

\usepackage{graphicx}
\usepackage{amsmath}
\usepackage{amssymb}
\usepackage{bm}
\usepackage{xcolor}
\usepackage[
  colorlinks=true,
  linkcolor=blue,
  citecolor=blue,
  urlcolor=blue,
  bookmarksnumbered=true,
  bookmarksopen=true
]{hyperref}

\graphicspath{{figures/}}

\begin{document}

\title{Rotation overcharges vectorized black holes}

\author{Shengjun Guo}
\affiliation{Department of Physics and Siyuan Laboratory, Jinan University,
             Guangzhou 510632, China}

\author{Guangzai Ye}
\affiliation{Institut de F\'{i}sica d'Altes Energies (IFAE),
             The Barcelona Institute of Science and Technology,
             Campus UAB, 08193 Bellaterra (Barcelona), Spain}
\affiliation{Port d'Informaci\'{o} Cient\'{i}fica (PIC),
             Campus UAB, C. Albareda s/n,
             08193 Bellaterra (Barcelona), Spain}

\author{Peng Liu}
\email[Corresponding author: ]{phylp@email.jnu.edu.cn}
\affiliation{Department of Physics and Siyuan Laboratory, Jinan University,
             Guangzhou 510632, China}

\date{\today}

\begin{abstract}
A charged black hole has a ceiling: extremality bounds the charge that a given mass and spin can support. 
Vector hair has so far only lowered it. In the Einstein--Maxwell--Vector (EMV) model, where a massless vector 
field couples nonminimally to the electromagnetic invariant, the spherically symmetric vectorized 
Reissner--Nordstr\"om black holes are strictly undercharged: vector hair is paid for in charge. Whether this ceiling 
is a property of the vector field or merely of spherical symmetry has remained open, because spherical symmetry 
hides the hair's mass as well: its Komar mass vanishes identically and the exact identity $M_V=2\Omega_HJ_V$ 
degenerates to $0=0$. Rotation lifts both degeneracies at once. We construct the rotating solutions of the model, 
the first in this setting, and find that the hair's mass is not a free parameter but a second-order invariant of 
the rotational response, fixed by $M_V=2\Omega_HJ_V$. Above a coupling threshold the ceiling gives way: the domain 
of existence crosses the extremal Kerr--Newman bound, reaching $q=Q/M>1$ with $q^2+\chi^2$ up to $1.84$, while 
at $\alpha=5$ the maximum charge stays below it. The shadow is where the hair becomes visible. 
At $(q,\chi)=(0.9947,0.10)$, near the extremal bound, the hair contributes an ellipticity of $8.30\%$, 
against $0.44\%$ from the charge and $0.06\%$ from rotation --- a factor of $19$ that charging a Kerr black hole 
cannot mimic; at the overcharged solution $(q,\chi)=(1.2527,0.10)$ the total ellipticity reaches $25.6\%$. 
Whether the branch is ghost-free remains open, but the existing no-go criteria do not settle it: they assume a 
background and a dispersion relation that do not hold here.
\end{abstract}

\maketitle

\section{Introduction}

A charged black hole has a ceiling. For Reissner--Nordstr\"om and Kerr--Newman (KN) solutions,
extremality bounds the charge that a given mass and spin can support, and this bound governs both
the causal structure of the exterior and the thermodynamics of the hole. Hair is usually expected to
relax such restrictions. For the vector hair of the Einstein--Maxwell--Vector (EMV) model the
opposite happens. Because the vector field couples nonminimally to the Maxwell invariant, its hair is
sourced by the very field that carries the charge, and the spherically symmetric vectorized
Reissner--Nordstr\"om black holes are \emph{strictly undercharged}~\cite{Oliveira:2020dru,Ye:2024pyy}:
the combined charge $\sqrt{\tilde Q^2+\tilde P^2}$, built from the electric and vector charges in
units of the ADM mass, is capped at unity and the existence line never reaches extremality. Vector hair, in spherical symmetry, is paid for in charge.

That leaves an open question with a sharp physical handle: is the ceiling a property of the vector
field, or of spherical symmetry? Spherical symmetry cannot answer it. It degenerates all three of the
hair's signatures at once --- the Komar mass of the vector field vanishes identically, the horizon
angular velocity vanishes so that the exact identity $M_V=2\Omega_HJ_V$ reduces to $0=0$, and the
charge is capped from above --- so that nothing distinguishes ``the vector field does not want more
charge'' from ``spherical symmetry does not let it have more''. Rotation lifts all three degeneracies
simultaneously, which is why it enters this paper not as one more parameter but as the symmetry
breaking that makes the vector field's mass and charge content visible.

None of this explains \emph{why} the hair exists; that is forced by the coupling rather than assumed. The function
$U(|\mathcal{B}|^2)=e^{\alpha|\mathcal{B}|^2}$ multiplies the Maxwell invariant, so that the
vector field acquires a background-dependent effective mass,
\begin{equation}
  m_{\rm eff}^2=\tfrac12\alpha\,e^{\alpha|\mathcal{B}|^2}F_{ab}F^{ab},
\end{equation}
which is negative whenever the electromagnetic field is electrically dominated. The hairless
configuration is then unstable, and the field grows until nonlinearities balance it: spontaneous
vectorization, the vector analogue of spontaneous scalarization in Einstein--Maxwell--scalar
theories~\cite{Herdeiro:2018wub,Doneva:2022ewd}. Hairy black holes are by now familiar objects: the
classical uniqueness theorems~\cite{Chrusciel:2012jk} are evaded both by nonminimal interactions and
by fields that do not inherit the spacetime symmetries~\cite{Herdeiro:2015waa,Bizon:1990sr}, and
explicit static and rotating constructions exist for scalar and Proca
fields~\cite{Herdeiro:2014goa,Herdeiro:2016tmi,Kleihaus:2015aje,Antoniou:2017acq,Burrage:2023zvk},
with curvature- and Maxwell-induced scalarization as the best-understood
mechanisms~\cite{Doneva:2017bvd,Silva:2017uqg,Herdeiro:2020wei,Cunha:2019dwb,Guo:2023mda,Xiong:2023bpl,Cheng:2025hdw}.

What is specific to the vector case is that the mechanism is currently under scrutiny, which is a
second reason to ask what its solutions look like. It has been argued that the tachyonic drive is
accompanied by ghost- or gradient-type instabilities~\cite{Pizzuti:2023eyt,Clough:2022ghost}, and a
recent no-go theorem shows that for perturbations of a hairless background the absence of ghosts and
gradients implies the absence of the tachyonic mode itself, forbidding a clean spontaneous
growth~\cite{Chiang:2026nogo}. Those results constrain \emph{how} vectorized black holes may form;
they do not by themselves determine \emph{what} stationary vectorized solutions exist, and we return
to the distinction in Sec.~\ref{sec:Conclusion}. Rotating vectorized black holes have recently been
constructed in Einstein-vector-Gauss-Bonnet (EvGB) theory~\cite{Kleihaus:2026evgb}, where the
coupling is to the Gauss--Bonnet invariant rather than to the Maxwell field. The two settings probe
complementary physics: because EvGB carries no Maxwell charge, its thermodynamic identity acquires no
$Q\Phi$ term and its solutions have no analogue of the undercharge ceiling, so that the reversal
reported here is structurally unavailable there. In the EMV model itself only spherical solutions are
known; the rotating sector, in which the ceiling and the two degeneracies are lifted, has not been
explored.

Beyond their theoretical interest, hairy black holes offer distinctive optical
signatures~\cite{Cunha:2018acu,Perlick:2021aok}. Ray-tracing studies have found strongly deformed or
disconnected shadows and chaotic lensing in selected scalar-haired
spacetimes~\cite{Cunha:2015yba,Cunha:2016bjh}, while fundamental photon orbits help explain the
shadow structure of more general rotating geometries~\cite{Cunha:2017eoe}. These effects are not
universal: some rotating scalarized solutions retain nearly Kerr-like
shadows~\cite{Fernandes:2024ztk}, and existing EHT bounds on scalar hair remain model
dependent~\cite{Cunha:2019ikd}. Studies of EMS black holes have also revealed enriched photon-ring and
higher-order image structures associated with modified photon-orbit
structures~\cite{Gan:2021xdl,Chen:2026rhr}, complementing broader work on photon rings as probes of
spacetime geometry~\cite{Johnson:2019ljv,Staelens:2023jgr}. Imaging has already constrained departures
from the Kerr description~\cite{EventHorizonTelescope:2019ggy,EventHorizonTelescope:2022xqj}, and
next-generation observations~\cite{Johnson:2023ynn} are expected to sharpen these tests.

Against this background we construct the rotating vectorized black holes of the EMV model, the first
in this setting. Rotation first makes the hair's mass content visible. It endows the hair with a
nonzero Komar mass contribution, which vanishes identically in the spherically symmetric limit and
scales approximately as $\chi^2$ for $\alpha=5$--$15$ in the low-spin regime; the analytic identity
$M_V=2\Omega_HJ_V$ is numerically verified to an accuracy of $\sim10^{-10}$ once the angular-grid
aliasing is eliminated, and it fixes the leading-order coefficient in terms of the linear rotational
responses, which is why no independent vector-hair contribution appears in the Smarr relation.

Having made that content visible, rotation also displaces the ceiling rather than respecting it. The domain of
existence crosses the extremal KN bound, extending to $q\approx1.35$; $q^2+\chi^2$ reaches $1.84$, and
$18.1\%$ of the explored solutions lie beyond the bound. The crossing happens only above
a coupling threshold: at $\alpha=5$ the maximum charge remains below the bound.

Observation is where the hair becomes visible in practice, and shadows are the natural place to
look, because their ellipticity is driven by the hair rather than by the charge. At
$(q,\chi)=(0.9947,0.10)$, near the extremal bound, the hair contributes
$8.30\%$, against $0.44\%$ from the charge and $0.06\%$ from rotation --- a factor of $19$ that no
model which merely charges a Kerr black hole can reproduce --- and at the overcharged solution
$(q,\chi)=(1.2527,0.10)$ the total ellipticity reaches $25.6\%$. Current imaging does not
yet constrain the quantity we compute: the EHT imaging parameters cannot recover ring ellipticity, and
axis ratios as extreme as $2:1$ remain consistent with the M87* reconstructions~\cite{Tiede:2022ell}. The ellipticity predicted below is therefore neither tested nor
excluded by existing data; it is a target for next-generation observations.

The remainder of this paper is organized as follows. Section~\ref{sec:EMVMODEL} introduces the EMV
model, numerical method, and physical quantities. Section~\ref{sec:Results} presents our main results
and some discussion, followed by our conclusions in Sec.~\ref{sec:Conclusion}. Throughout this paper,
we adopt geometrized units in which $G=c=4\pi\epsilon_0=1$.
\section{The Einstein--Maxwell--Vector Model}
\label{sec:EMVMODEL}

\subsection{The Action, Equations of Motion, and Ansatz}
\label{subsec:Eqsofmotion}

The Einstein--Maxwell--Vector model in spherical symmetry has been 
studied extensively in Refs.~\cite{Oliveira:2020dru,Ye:2024pyy}. 
We begin by introducing the basic setup. The action of the model is given by
\begin{equation}
  S = \frac{1}{16\pi} \int d^4 x \sqrt{-\mathrm{g}} \left[ R - 
      U\left(|\mathcal{B}|^2\right) F^{ab} F_{ab} - V^{ab} V_{ab} \right],
\end{equation}
where $R$ is the Ricci scalar, $F_{ab}$ is the electromagnetic 
field tensor corresponding to the 4-potential $\mathcal{A}^a$. 
$V_{ab}$ denotes the field-strength tensor associated with a massless vector 
field characterized by $\mathcal{B}^a$, which is defined as 
$V_{ab}=\nabla_a \mathcal{B}_b-\nabla_b \mathcal{B}_a$. Here $U$ denotes the coupling function 
between the electromagnetic field and the vector field. We choose an exponential form for this function, 
namely
\begin{equation}
\label{couple}
  U\left( |\mathcal{B}|^2 \right) = e ^{\alpha |\mathcal{B}|^2},
\end{equation}
where $\alpha$ is a coupling constant.

By varying the action with respect to the metric, the electromagnetic 
field, and the vector field, we obtain their respective equations of 
motion
\begin{align}
  &R_{ab} - \frac{1}{2} R g_{ab} = 2 \left(T^F_{ab} + T^V_{ab}\right), \label{eommetric} \\
  &\nabla_a \left( U\left(|\mathcal{B}|^2\right) F^{ab} \right) = 0, \label{eomF} \\
  &\nabla_a V^{ab} = \frac{1}{2} U'\left(|\mathcal{B}|^2\right) F^{cd} F_{cd} \mathcal{B}^b, \label{eomV}
\end{align}
where $T^F_{ab}$ and $T^V_{ab}$ are the energy-momentum tensors of the
electromagnetic field and the vector field, respectively
\begin{align}
  T_{ab}^V &\equiv V_a{}^c V_{bc} - \frac{1}{4} V_{cd} V^{cd} g_{ab} + 
                   \frac{1}{2} U'\left(|\mathcal{B}|^2\right) F_{cd} F^{cd} \mathcal{B}_a \mathcal{B}_b, \label{eq:defTV} \\
  T_{ab}^F &\equiv U\left(|\mathcal{B}|^2\right) \left( F_a{}^c F_{bc} - \frac{1}{4} 
                   F_{cd} F^{cd} g_{ab} \right).
\end{align}
Eq.~\eqref{eomV} can be regarded as a Proca-like equation. Substituting 
Eq.~\eqref{couple} into Eq.~\eqref{eomV}, we obtain
\begin{equation}
\label{effmass}
  \nabla_a V^{ab} = \frac{1}{2} \alpha U\left(|\mathcal{B}|^2\right) F^{cd} F_{cd} \mathcal{B}^b 
                  = \mu_{\text{eff}}^2 \mathcal{B}^b.
\end{equation}
On the Reissner--Nordström black hole background, $F^{cd}F_{cd}<0$. 
Therefore, for $\alpha>0$, the squared effective mass of 
the vector field becomes negative, triggering a tachyonic 
instability. Consequently, the vector field grows 
exponentially, leading to the spontaneous vectorization of 
the RN black hole.

In this work, we focus on stationary, asymptotically flat and axisymmetric 
black hole solutions with vector hair, described by the equations 
of motion \eqref{eommetric}, \eqref{eomF} and \eqref{eomV}. As such, 
the following metric ansatz is reasonable to consider
\begin{equation}
\label{metricansatz}
  ds^2 = -f N^2 dt^2 + \frac{g}{f} 
                  \left[ h \left( dr^2 + r^2 d\theta^2 \right) 
                        + r^2 \sin^2\theta \left( d\varphi - \frac{W}{r}\left(1 - N\right) dt \right)^2 
                  \right],
\end{equation}
where $N = 1 - r_H/r$, and $f, g, h, W$ are functions 
of $r$ and $\theta$, $r_H$ represents the event horizon radius.
The coordinates used here are quasi-isotropic coordinates, 
introduced systematically in Ref.~\cite{Fernandes:2022gde}. 
Their only difference from the Boyer--Lindquist coordinates lies in the radial coordinate $r$.
For the electromagnetic four-potential and the vector field, the ansatz is given by
\begin{align}
  \mathcal{A}_\mu dx^\mu 
    &= \left( A_t - \frac{W}{r}\left(1 - N\right) A_\varphi \sin^2\theta \right) dt 
      + A_\varphi \sin^2\theta d\varphi, \\
  \mathcal{B}_\mu dx^\mu 
    &= \left( B_t - \frac{W}{r}\left(1 - N\right) B_\varphi \sin^2\theta \right) dt 
      + B_\varphi \sin^2\theta d\varphi,
\end{align}
where $A_t$, $A_\varphi$, $B_t$ and $B_\varphi$ are functions of $r$ and $\theta$.
The electromagnetic four-potential $\mathcal{A}_\mu$ is not uniquely determined, owing to its gauge freedom. We choose 
$\left.\mathcal{A}_\mu \xi^\mu\right|_{H}=0$, where $\xi^a = t^a + \Omega_H \varphi^a$ 
and $\Omega_H$ is the angular velocity of the horizon, 
and will specify this gauge condition explicitly when imposing the 
boundary conditions below.

\subsection{Numerical Implementation}

To solve this highly nonlinear system of partial differential equations (PDEs), we simplify the 
metric equations. Writing $E_{ab} = R_{ab} - \frac{1}{2} g_{ab}R - 2T_{ab}$, we recast them so that 
each equation contains second-order derivatives of a single function. This is achieved by taking 
the following combinations:
\begin{align}
  {E^t}_t - {E^r}_r - {E^\theta}_\theta - {E^\varphi}_\varphi + \frac{2r_H W}{r^2} {E^t}_\varphi &= 0, \\
  {E^\varphi}_\varphi - {E^r}_r - {E^\theta}_\theta - \frac{r_H W}{r^2} {E^t}_\varphi &= 0, \\
  {E^r}_r + {E^\theta}_\theta &= 0, \label{Prrg} \\
  {E^t}_\varphi &= 0. 
\end{align}
Together with Eqs.~\eqref{eomF} and \eqref{eomV}, each of which consists of 
two component equations, these equations form a system of PDEs involving eight 
unknown functions $B_t, B_\varphi, A_t, A_\varphi, f, g, h$ and $W$.

We now impose appropriate boundary conditions on this system. 
First of all, at spatial infinity, the spacetime is required to be asymptotically flat, 
while the vector fields vanish. Accordingly, we impose $\left(r \to \infty\right)$
\begin{equation}
\begin{aligned}
  & f = g = h = 1, \quad \frac{r^2}{2r_H} \partial_r W + \left( 1 + \frac{r^2}{2r_H} \partial_r f \right)^2 \chi = 0, \\
  & \frac{r^2}{2r_H} \partial_r A_t - \frac{q}{2} \left( 1 + \frac{r^2}{2r_H} \partial_r f \right) = 0, \quad A_\varphi = 0, \quad B_t = B_\varphi = 0,
\end{aligned}
\end{equation}
where $\chi \equiv J/M^2,\ q \equiv Q/M$, with $M, J, Q$ being the ADM mass, angular momentum, and electric charge respectively. 
Furthermore, at the event horizon, we require the metric functions to be regular $\left(r=r_H\right)$,
\begin{equation}
\begin{aligned}
  & f - r_H \partial_r f = 0, \quad g + r_H \partial_r g = 0, \quad \partial_r h = 0, \quad W - \frac{r_H}{2} \partial_r W = 0, \\
  & A_t = 0, \quad \partial_r A_\varphi = 0, \quad B_t = 0, \quad \partial_r B_\varphi = 0.
\end{aligned}
\end{equation}
In the vicinity of the symmetry axis, all functions are required to be smooth. 
To avoid conical singularities, we impose
\begin{equation}
\begin{aligned}
  & \partial_\theta f = \partial_\theta g = \partial_\theta h = \partial_\theta W = \partial_\theta A_t = \partial_\theta A_\varphi = \partial_\theta B_t = \partial_\theta B_\varphi = 0, \\ 
  & \text{and} \quad h = 1, \quad \text{for} \quad \theta = 0,\pi.
\end{aligned}
\end{equation}
Finally, since the system is symmetric with respect to the equatorial plane, 
the boundary conditions at the equatorial plane are given by
\begin{equation}
  \partial_\theta f = \partial_\theta g = \partial_\theta h = \partial_\theta W = \partial_\theta A_t = \partial_\theta A_\varphi = \partial_\theta B_t = \partial_\theta B_\varphi = 0, \quad 
  \text{for} \quad \theta = \pi/2.
\end{equation}

To solve the equations, we introduce the coordinate transformation
\begin{equation}
  x=1-\frac{2r_H}{r},
\end{equation}
which maps $r\in[r_H,+\infty)$ to $x\in[-1,1]$.
We employ a pseudospectral method combined with the Newton--Raphson root-finding 
method to solve the nonlinear system of PDEs numerically. We expand each function spectrally in 
the radial coordinate $x$ and the angular coordinate $\theta$, with resolutions $n_x$ and $n_\theta$, 
respectively. Specifically, each function $\mathcal{F}^{(k)} \in \{B_t,B_\varphi,A_t,A_\varphi,f,g,h,W\}$ 
is represented in the following form:
\begin{equation}
\mathcal{F}^{(k)}=\sum_{i=0}^{n_x-1}\sum_{j=0}^{n_\theta-1}c_{ij}^{(k)}T_i(x)\cos(2j\theta),
\end{equation}
where $T_i(x)$ denotes the Chebyshev polynomial of the first kind of degree $i$. 
For a detailed discussion of the method we refer the reader to Refs.~\cite{Fernandes:2022gde,Burrage:2023zvk}. 
Since this method is highly sensitive to the initial guess, we use the spherically 
symmetric numerical solutions reported in \cite{Ye:2024pyy} as initial guesses to obtain a numerical solution 
with a small amount of rotation. By subsequently varying the parameters 
incrementally, we obtain solutions for arbitrary parameter values, provided 
that a solution exists there.
Throughout this study we fix $r_H=1$, leaving the three parameters $\{q,\chi,\alpha\}$ 
that can be varied. Unless stated otherwise, we use a grid resolution of 
$n_x \times n_\theta = 60\times20$.

Before closing this subsection, we return to Eq.~\eqref{Prrg}. This equation is a relatively simple 
second-order partial differential equation involving only the function $g$, and under the boundary 
conditions imposed above it admits the analytical solution
\begin{equation}
  \label{eq:fung}
  g=\left(1+\frac{r_H}{r}\right)^2,
\end{equation}
in agreement with previous results \cite{Ye:2024pyy,Burrage:2023zvk}. In the present study, however, we substitute this 
analytic expression into the PDEs system only when higher numerical accuracy is required. The
field equations used by the solver, their independent symbolic re-derivation, and the verification
suite that checks the archived solutions against them are collected in Ref.~\cite{Guo:2026repro}.

\subsection{Physical Quantities}

For the spacetime considered in this work, many physical quantities can be 
extracted from the asymptotic behavior of the metric components and the 
corresponding functions. At spatial infinity, we have
\begin{equation}
  g_{tt} \sim -1 + \frac{2M}{r}, \quad 
  g_{\varphi t} \sim -\frac{2J}{r} \sin^2\theta, \quad 
  A_t \sim \Phi - \frac{Q}{r}, \quad
  B_t \sim \frac{P}{r},
\end{equation}
where $\Phi$ denotes the difference in the electric potential between 
spatial infinity and the event horizon. Next we consider the horizon quantities; here and below a subscript $H$ denotes 
evaluation at the event horizon. The angular velocity of the horizon is
\begin{equation}
  \Omega_H = \left. W / r_H \right|_{r_H}.
\end{equation}
 The Hawking temperature $T_H$ and the horizon area $A_H$ are given by
\begin{align}
  T_H = \frac{1}{2\pi r_H} \left. \frac{f}{\sqrt{gh}} \right|_{r_H}, \quad
  A_H = 2\pi {r_H}^2 \left. \int_0^\pi d\theta \sin\theta \frac{g\sqrt{h}}{f} \right|_{r_H}.
\end{align}
Since no curvature-induced modified-gravity terms are included in the model, 
the black hole entropy is $S = A_H/4$. The quantities introduced above can be 
related through a Smarr relation as follows:
\begin{equation}
  M = 2 T_H S + 2 \Omega_H J + Q \Phi. \label{SmarrnoV}
\end{equation}
This relation appears to be identical to the Smarr formula for Kerr--Newman
black holes (KNBHs). In Appendix~\ref{app:Smarr}, we provide a rigorous proof of 
Eq.~\eqref{SmarrnoV} within the framework of formal theory \cite{Wald:1993nt,Iyer:1994ys,Liberati:2015xcp}.
Accordingly, the first law governing rotating EMV black holes can be written as
\begin{equation}
  \label{eq:firstlaw}
  dM = T_H dS + \Omega_H dJ + \Phi dQ.
\end{equation}

Additionally, we employ the volume-integral form of the Komar formula \cite{waldGeneralRelativity} 
to compute the contributions of the vector field to the mass and angular momentum:
\begin{align}
  M_V &= \frac{1}{4\pi} \int_\Sigma  \left( 2T^V_{ab} - T^V g_{ab} \right) t^a d\Sigma^{b}, \label{MV} \\
  J_V &= -\frac{1}{8\pi} \int_\Sigma  \left( 2T^V_{ab} - T^V g_{ab} \right) \varphi^a d\Sigma^{b}. \label{JV}
\end{align}
Here $T^V_{ab}$ is the energy--momentum tensor of the vector field, defined in Eq.~\eqref{eq:defTV}, and
$T^V\equiv g^{ab}T^V_{ab}$ its trace.

\section{Results and Discussion}
\label{sec:Results}

\subsection{Mass of the Vector Field}
\label{subsec:massofV}

\begin{figure}[!htbp]
    \centering
    \includegraphics[width=5.355in]{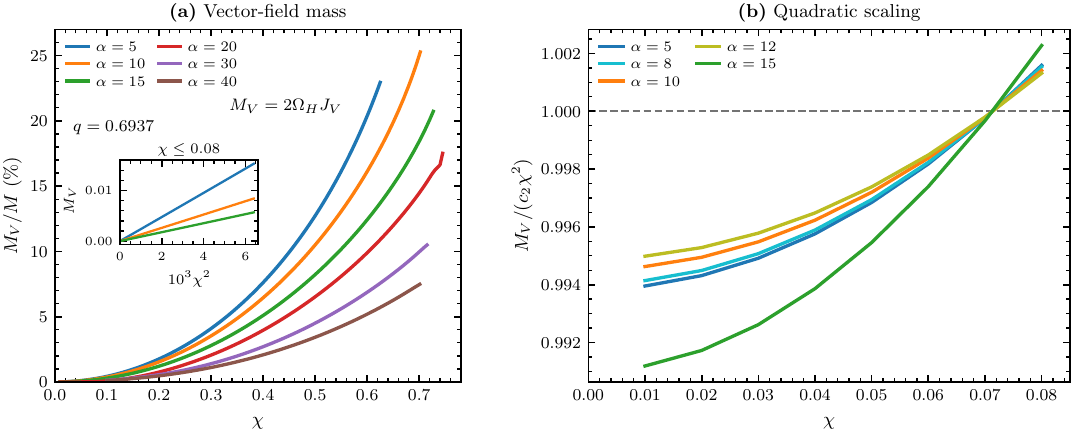}
    \caption{
    Vector-field Komar mass and its low-spin scaling at fixed
    $q=0.6937$.
    \textbf{(a)} Fraction of the ADM mass contributed by the vector
    field $\mathcal{B}$, $M_V/M$, expressed as a percentage,
    versus the dimensionless spin $\chi$ for
    $\alpha=5,10,15,20,30,40$.
    The inset shows $M_V$ versus $\chi^2$ for $\alpha=5,10,15$
    over $0.01\leq\chi\leq0.08$, with the horizontal coordinate
    displayed as $10^3\chi^2$.
    \textbf{(b)} Normalized mass $M_V/(c_2\chi^2)$ for
    $\alpha=5,8,10,12,15$ over the same low-spin interval, where the normalization
    $c_2$ is the coefficient obtained by a $\chi^2$-weighted least-squares fit of
    $M_V$ to $c_2\chi^2$ over that same interval. The panel therefore shows the
    quality of that fit, not an independent test of the quadratic law.
    The horizontal dashed line denotes exact quadratic scaling.
    Colors identify the coupling constants according to the
    legend in each panel.
    }
    \label{fig:MassofV}
\end{figure}

In the static, spherically symmetric case, Ref.~\cite{Ye:2024pyy} showed that the Komar mass contributions from the different 
components of the vector field exactly cancel, yielding $M_V=0$. Once rotation is introduced, however, this cancellation 
is broken: we find that rotation qualitatively changes the role of the vector hair in the mass decomposition of the black hole, 
and the vector hair acquires a nonzero contribution to the Komar mass. Figure~\ref{fig:MassofV}(a) shows the fraction of the total ADM mass
carried by the vector field, $M_V/M$, expressed as a percentage,
as a function of the dimensionless spin $\chi$ at fixed $q=0.6937$.
The six curves correspond to $\alpha=5,10,15,20,30,40$, and $M_V$
is computed from Eq.~\eqref{MV}.
The fractional contribution increases with spin along each sequence.
At fixed $\chi$ within the common range of the sequences, increasing
$\alpha$ reduces $M_V/M$.

The Komar mass of the hair is not an independent quantity. For every solution constructed here it satisfies the exact
relation
\begin{equation}
  M_V = 2\Omega_H J_V\,,
\end{equation}
derived below from the Komar identities together with the Smarr relation, and this relation has a sharp numerical
consequence. In the low-spin limit, both $\Omega_H$ and $J_V$ are linear in
$\chi$ at leading order, so that $M_V$ begins at quadratic order.
Writing $\Omega_H=\omega_1\chi+O(\chi^3)$ and
$J_V=j_1\chi+O(\chi^3)$ gives $M_V=2\omega_1j_1\chi^2+O(\chi^4)$.
Fitting $M_V\propto\chi^{p}$ over $0.01\leq\chi\leq0.08$ gives $p=2.003$--$2.005$ for
$\alpha=5$, $8$, $10$, $12$ and $15$, consistent with the approximately linear dependence of
$M_V$ on $\chi^2$ shown in the inset of Fig.~\ref{fig:MassofV}(a). We stress that this is a
\emph{corollary} of the identity $M_V=2\Omega_HJ_V$ together with the linearity of
$\Omega_H(\chi)$ and $J_V(\chi)$, not an independent measurement: it follows algebraically
from the expansion displayed above. The exponent is quoted only for $\alpha=5$--$15$; for
$\alpha=20$ the same fit gives $p\simeq2.28$--$3.24$, the deviation arising from the
threshold behaviour of the existence domain discussed in Sec.~\ref{sec:Results}.
Figure~\ref{fig:MassofV}(b) shows the departure of the vector-field mass
from a purely quadratic dependence over $0.01\leq\chi\leq0.08$.
For all five sequences, the ratio $M_V/(c_2\chi^2)$ deviates from unity
by no more than approximately $1\%$. Here $c_2$ is the fitted coefficient defined in the
caption of Fig.~\ref{fig:MassofV}, so this statement quantifies the goodness of the
quadratic fit; an independent check is provided below by comparing $c_2$ with the
analytic value $2\omega_1j_1$.
These small departures from the horizontal reference line indicate the
accuracy of a purely quadratic description over this finite spin interval.
The linear term is bounded by $|c_1|/c_2\lesssim3\times10^{-6}$, which is the statement that
parity forbids a term linear in the spin. We do not quote a value for $M_V(\chi\to0)$: a
constant term can only be obtained by fitting a polynomial that includes one, and the result
is not statistically significant, changes by one to two orders of magnitude with the fit
window, and changes sign across the sequences, so it is not a meaningful quantity. The
quadratic coefficient is likewise fixed: it agrees with $2\omega_1j_1$, where
$\omega_1$ and $j_1$ are the linear rotational responses of $\Omega_H$ and $J_V$, to within $0.5\%$. The vector mass is
therefore a \emph{second-order invariant of the rotational response} rather than a new degree of freedom appearing at
first order in the spin: it vanishes identically on spherically symmetric configurations that do carry vector hair, and
it is completely determined by the linear response of the rotating solution.

This structure is also visible directly in the integrand of the Komar volume integral. Evaluating the contribution of
the vector field to $M_V$ term by term, every term contains at least two factors drawn from
$\{B_t,B_\varphi,W\}$; since $B_\varphi$ and $W$ both vanish in the spherically symmetric limit while $B_t$ does not,
the vanishing of $M_V$ in that limit is manifest, and the leading rotational contribution is necessarily of second
order. Equation~\eqref{MV} combined with this counting explains both the measured exponent and the fact that the
fractional contribution $M_V/M$ decreases with increasing $\alpha$: a larger coupling produces a smaller
$B_\varphi$ response for fixed spin.

However, $M_V$ does not appear as an independent term in Eq.~\eqref{SmarrnoV}. In other words, rotation 
allows the vector hair to make a nonzero contribution to the Komar mass, but this contribution is encoded in 
the global quantities characterizing the spacetime rather than entering the Smarr relation as an additional term.

The vector hair does not carry an independent global conserved charge analogous to the Maxwell electric charge, 
and thus the absence of an explicit $M_V$ term in Eq.~\eqref{SmarrnoV} is not in itself contradictory. Nevertheless, 
this does not preclude us from further exploring this feature. Starting from the surface and volume integrals 
in the Komar identity, with the surface integral evaluated at spatial infinity and on the event horizon, 
and the volume integral taken over the entire region exterior to the horizon, we obtain
\begin{equation}
  \oint_{\partial \Sigma} dS_{ab} \left( -\frac{1}{8\pi} \nabla^a \xi^b \right)
  =
  \int_\Sigma d\Sigma^b \left( \frac{1}{4\pi} \left(2T_{ab}-Tg_{ab}\right)\xi^a \right),
\end{equation}
which yields the following relation:
\begin{equation}
  M = 2T_H S + 2\Omega_H J + Q\Phi + M_V - 2\Omega_H J_V.
\end{equation}
This provides the usual route to the Smarr relation. Comparing with Eq.~\eqref{SmarrnoV}, we find that the vector 
hair must satisfy
\begin{equation}
  M_V = 2\Omega_H J_V.
\end{equation}
This relation indicates that the nonvanishing Komar mass of the vector hair is directly tied to rotation: 
the vector field contributes to the total mass only when the black hole possesses angular momentum, 
while its contribution vanishes in the spherically symmetric limit.

\subsection{Domain of Existence}

\begin{figure}[!htbp]
    \centering
    \includegraphics[width=5.95in]{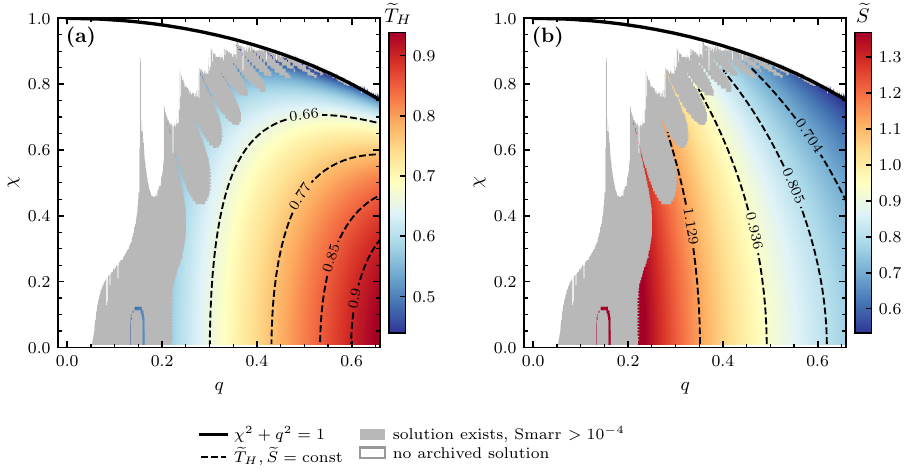}
    \caption{Domain of existence of rEMVBHs in the $q$--$\chi$ parameter space at $\alpha=20$ 
             for $q<0.65$, with $q$ on the horizontal axis and $\chi$ on the vertical axis. The color scale represents 
             the values of $\widetilde T_H$ {\bfseries (a)} and $\widetilde S$ {\bfseries (b)}. The black solid line denotes the reference 
             boundary of extremal KNBHs, $\chi^2+q^2=1$, while the dashed lines indicate contours of 
             constant $\widetilde T_H$ {\bfseries (a)} and $\widetilde S$ {\bfseries (b)}. }
    \label{fig:lowq}
\end{figure}

\begin{figure}[!htbp]
    \centering
    \includegraphics[width=5.95in]{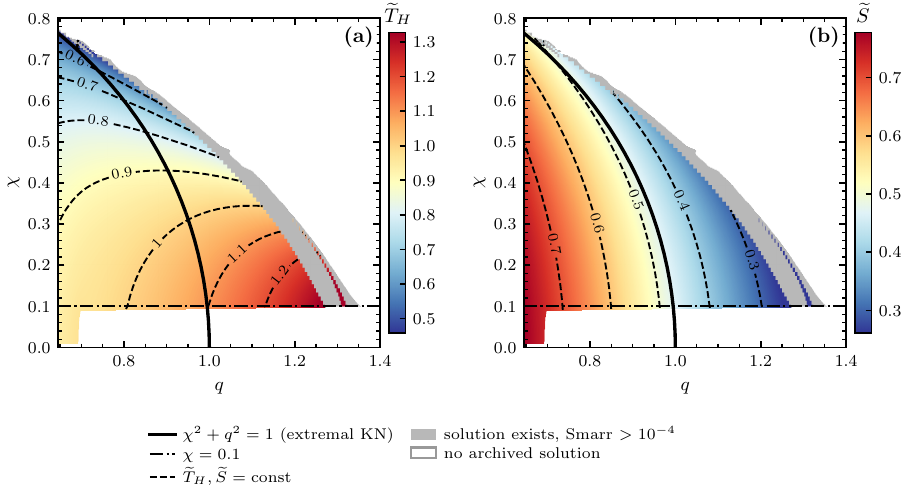}
        \caption{Domain of existence of rEMVBHs for $\alpha=20$ and $q>0.65$, with $q$ on the horizontal axis and
             $\chi$ on the vertical axis. The black solid line denotes the reference boundary of extremal KNBHs,  
             $\chi^2+q^2=1$. As in Fig.~\ref{fig:lowq}, panels {\bfseries (a)} and {\bfseries (b)} show the color maps of 
             $\widetilde T_H$ and $\widetilde S$, respectively.}
    \label{fig:highq}
\end{figure}

We have explored the full set of numerical solutions in the $q$--$\chi$ parameter space at fixed coupling constant $\alpha=20$. 
Figures~\ref{fig:lowq} and \ref{fig:highq} show the domains of existence of rotating EMV black holes (rEMVBHs) for $q<0.65$ and 
$q>0.65$, respectively. Color maps are employed to illustrate the variation of $\widetilde T_H$ and $\widetilde S$ across 
the $q$--$\chi$ parameter space. The two regions are presented separately because different numerical continuation strategies 
are used. In Fig.~\ref{fig:lowq}, $q$ is held fixed while $\chi$ is increased to determine its upper bound, whereas in 
Fig.~\ref{fig:highq} the continuation is performed in the opposite direction.

The Hawking temperature and black hole entropy used here are reduced quantities. 
The EMV model possesses the scaling symmetry $r\to\lambda r$ and $Q\to\lambda Q$, where $\lambda>0$ is a constant. 
Under this transformation, the remaining physical quantities scale accordingly; for example, $M\to\lambda M$, 
while the coupling function $U$ remains invariant. Accordingly, we introduce the following reduced quantities to 
facilitate a clearer comparison of the physical observables across different regions of parameter space.
\begin{equation}
  \widetilde T_H = 8\pi M T_H,\quad \widetilde{S} = \frac{S}{4\pi M^2}.
\end{equation}

As shown in Fig.~\ref{fig:lowq}, in the low-charge regime the rEMVBHs remain entirely within the domain of existence of 
KNBHs. Rotation tends to reduce the allowed range of $q$, and this effect becomes more pronounced as 
$\chi$ increases. In panel {\bfseries (a)}, the $\widetilde T_H$ contours are ordered so that their values decrease 
from right to left. Thus, moving toward the left boundary of the existence domain corresponds to progressively lower 
Hawking temperatures. Since extremality is characterized by $\widetilde T_H\to0$, this behavior indicates that the numerical 
solutions are approaching an extremal rEMVBH configuration.

A qualitatively different behavior is observed in Fig.~\ref{fig:highq}, and it is in direct contrast with the
spherically symmetric case. There, the combined charge $\sqrt{\tilde Q^2+\tilde P^2}$ --- with $\tilde Q$ and
$\tilde P$ the electric charge and vector charge in units of the ADM mass, so that $\tilde Q=q$ for the purely electric
case --- obeys a ceiling at unity, 
so that vectorized Reissner--Nordstr\"om black holes remain undercharged and the associated existence line at
$\alpha=20$ terminates at $q_{\max}^{\rm sph}=0.6947$~\cite{Oliveira:2020dru,Ye:2024pyy}. The rotating families constructed here
begin at $q=0.6937$ and $0.6947$, i.e.\ within $0.15\%$ and $0.005\%$ of that existence line, respectively, so that the
rotating families originate on the spherical existence line; they then extend to $q$ as large as $1.35$, close to
twice the spherical ceiling and far above the extremal KN value at the same spin ($0.9939$ at $\chi=0.11$). Of the
$72\,480$ archived $\alpha=20$ solutions, $18.1\%$ lie beyond $q^2+\chi^2=1$, the largest excess being
$q^2+\chi^2=1.8425$ at $(q,\chi)=(1.3537,0.10)$. That extremal point lies at the edge of the
numerically converged region: its dimensionless Smarr residual is $2.7\times10^{-2}$, far above
the accuracy threshold $10^{-4}$ that defines the colour/grey partition of
Fig.~\ref{fig:highq}. Restricting to accuracy-qualified solutions (dimensionless residual
below $10^{-4}$, $56\,685$ of the $72\,480$), $17.6\%$ still lie beyond the bound and the
largest excess is $q^2+\chi^2=1.7542$. The per-point residual archive, together with the code
that recomputes every quantity quoted in this section directly from the raw solution files, is
available in the accompanying public repository~\cite{Guo:2026repro}, where a single command
re-executes the full verification suite. The qualitative conclusion is therefore unaffected:
a large fraction of \emph{converged} solutions still exceed the extremal Kerr--Newman
bound, so overcharging is not an artefact of the unconverged tail of the archive. Rotation therefore reverses the undercharging that holds in spherical
symmetry, and Fig.~\ref{fig:highq} shows that the reversal is thresholded in the coupling rather than generic: at
$\chi=0.30$ the largest charge that can be continued numerically is $q_{\max}^{\rm cont}=0.9427$ for $\alpha=5$, still
below the extremal KN value $0.9539$ (the spherical existence line at this coupling terminates even earlier, at
$q=0.938$), and only for $\alpha\gtrsim8$ does the domain of existence cross the bound. Around $\chi\simeq0.1$,
$q_{\max}^{\rm cont}$ increases sharply and extends beyond the extremal KN bound. As $\chi$ is increased further,
however, it gradually approaches the extremal KN boundary. We emphasise that three distinct quantities are involved and
should not be conflated: the endpoint of the spherical existence line, $q_{\max}^{\rm sph}(\alpha)$; the largest charge
that numerical continuation reaches, $q_{\max}^{\rm cont}(\alpha,\chi)$; and the edge of the accuracy-qualified
(colored) region, $q_{\rm acc}(\alpha,\chi)$, which at $\alpha=20$ lies $0.03$--$0.10$ lower in $q$ and marks those
solutions that satisfy the accuracy criteria discussed below.

Finally, both Figs.~\ref{fig:lowq} and \ref{fig:highq} contain gray regions that are not assigned any values of $\widetilde T_H$ 
or $\widetilde S$. These regions require clarification. In exploring the $q$--$\chi$ parameter space, we employed a grid 
resolution of $n_x\times n_\theta=60\times20$. At this resolution, the colored regions represent numerical solutions for which 
the residual of the Smarr relation in Eq.~\eqref{SmarrnoV} is below $10^{-4}$. In contrast, the gray regions correspond to solutions that 
fail to satisfy this accuracy criterion. 
This does not imply that numerical solutions cease to exist in these regions. Rather, we regard them as regions in which 
the numerical error is too large for the extracted physical quantities to be considered reliable at the adopted resolution. 
We have further computed a large number of solutions within the gray regions and found that increasing the grid resolution
and incorporating the analytic solution for $g$ given in Eq.~\ref{eq:fung} into the PDEs system can substantially
improve numerical accuracy, as evidenced by a significant reduction in the residual of the Smarr relation. Both the
residual data underlying this partition and the scripts that reproduce it from the archived solutions are provided in
Ref.~\cite{Guo:2026repro}.
Appendix~\ref{app:solplots} presents three representative numerical 
solutions located along the left boundary of the gray region in Fig.~\ref{fig:lowq}, computed using a $100\times50$ grid. 
The corresponding model parameters and physical quantities are summarized in Table~\ref{tab:3sols}.

\begin{table}[htbp]
  \centering
  \caption{Parameters and physical quantities of three solutions along the left
           boundary of the gray region in Fig.~\protect\ref{fig:lowq},
           computed with a $100\times50$ grid. 
           Here $\Delta_{\rm Smarr}\equiv|M-2T_HS-2\Omega_HJ-Q\Phi|/M$ denotes the dimensionless residual of the Smarr relation, the same quantity that defines the colour/grey partition of Fig.~\ref{fig:highq}.}

  \label{tab:3sols}
  \setlength{\tabcolsep}{2.5pt}
  \begin{tabular}{cccccccccc}
    \hline\hline
        Solution & $q$ & $\chi$ & $T_H$ & $A_H$ & $P$ & $M$ & $J$ & $\Omega_H$ & $\Delta_{\rm Smarr}$ \\
    \hline
        I   & 0.1977 & 0.515 & 0.0078442 & 509.93 & 1.3225 & 2.6971 & 3.7463 & 0.034358 & $6.3324\times10^{-7}$ \\
        II  & 0.1487 & 0.450 & 0.0073495 & 544.29 & 1.3084 & 2.6023 & 3.0474 & 0.030064 & $2.1773\times10^{-5}$ \\
        III & 0.1387 & 0.347 & 0.0074867 & 534.30 & 1.3109 & 2.5185 & 2.2010 & 0.023165 & $1.6555\times10^{-5}$ \\
    \hline\hline
  \end{tabular}
\end{table}

\subsection{Shadows}
\label{sec:Shadows}

The black hole shadow provides one of the most direct visual manifestations of a black hole spacetime. In recent years, 
substantial progress has been made in both the analytical and numerical study of black hole shadows across a wide variety 
of gravitational models; see Refs.~\cite{Perlick:2021aok,Cunha:2018acu,Gralla:2019xty} for reviews. 
In this subsection, we employ the open-source code FOORT (see Refs.~\cite{OSS:FOORT,Staelens:2023jgr,Fernandes:2024ztk}) 
to compute the shadow contours of rEMVBHs. The code implements a backward ray-tracing scheme, in which the initial 
four-momentum $p^\mu$ of each photon is constructed on the observer's screen and the corresponding past-directed null 
geodesic is integrated backward through the spacetime. Each ray is then classified according to whether it escapes to 
the exterior region or is captured by the black hole. Once the outcomes for all pixels on the observer's screen have 
been determined, the boundary of the black hole shadow can be extracted using an ordered contour-tracing procedure.

\begin{figure}[!htbp]
    \centering
    \includegraphics[width=5.355in]{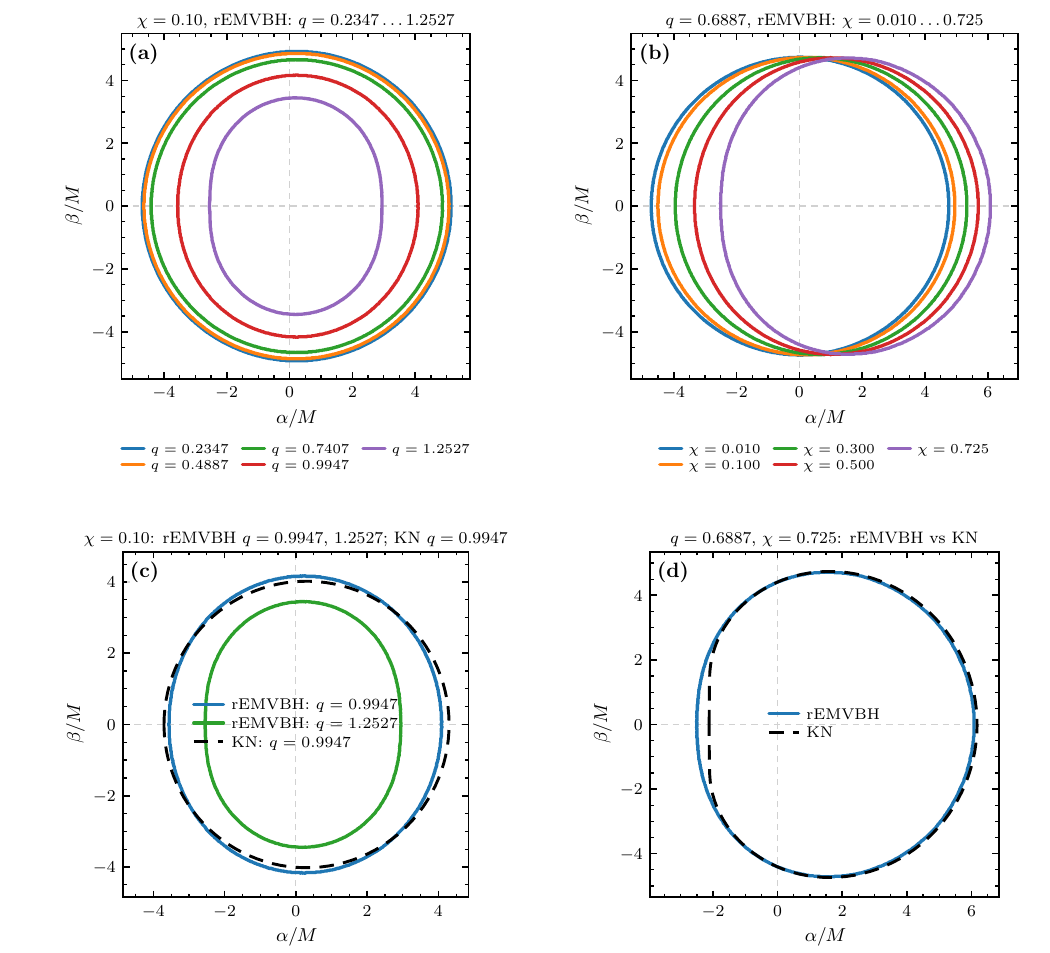}
    \caption{Shadow contours of rEMVBHs. Panels {\bfseries (a)} and {\bfseries (b)} show rEMVBHs with fixed $\chi=0.10$ while $q$ is varied ($q=0.2347$, $0.4887$, $0.7407$, $0.9947$, $1.2527$) and with fixed $q=0.6887$ while $\chi$ is varied ($\chi=0.010$, $0.100$, $0.300$, $0.500$, $0.725$), respectively. Panels {\bfseries (c)} and {\bfseries (d)} compare the shadows of rEMVBHs with those of KNBHs for selected near-extremal configurations: at $\chi=0.10$, the rEMVBH shadows with $q=0.9947$ and $q=1.2527$ against the KN shadow with $q=0.9947$, and at $(q,\chi)=(0.6887,0.725)$, a rEMVBH against its KN counterpart. Solid curves are obtained by ray tracing the numerical rEMVBH metrics with \textsc{foort},
    in the observer's local orthonormal frame at $r_{\rm obs}=1000\,M$; dashed curves are the analytic
    KN shadows, evaluated with the asymptotic Bardeen impact parameters $(\alpha,\beta)$. Applied to a
    KN metric, the same pipeline reproduces the analytic KN shadow to
    $|\Delta W/W|,\,|\Delta H/H|\lesssim0.1\%$ in width and height, and to $\lesssim0.02\,M$ in contour
    position.}
    \label{fig:Shadows}
\end{figure}

Throughout this work, the observer is placed at $r_{\rm obs}=1000M$ and $\theta_{\rm obs}=\pi/2$, since the 
displacement of the shadow is generally most pronounced for an equatorial observer. The physical size of the observer's 
screen is fixed to $15M\times15M$. Figure~\ref{fig:Shadows} presents the shadow contours of representative rEMVBHs 
at $\alpha=20$. Note that the screen coordinates $\alpha/M$ and $\beta/M$ are given in reduced form. The solid curves correspond 
to rEMVBHs, whereas the dashed curves represent the shadow contours of KNBHs. The latter are obtained 
from the corresponding analytic expressions; see Refs.~\cite{Tsukamoto:2017fxq,Shaikh:2019fpu}. The analytic form of the 
KN metric in the quasi-isotropic coordinates adopted in this work is given in the appendix of Ref.~\cite{Burrage:2023zvk}.

Before analyzing the shadow properties of rEMVBHs, we briefly review the characteristic behavior of 
KN black hole shadows, studied previously in Ref.~\cite{deVries:1999tiy}. (i) For small spin 
$\chi$, the KN shadow remains nearly circular, resembling that of a Reissner--Nordström black hole. 
As $q$ increases, the overall size of the shadow decreases, while its shape remains nearly unchanged. (ii) As $\chi$ 
increases, the shadow is progressively displaced in the direction of rotation, with the displacement becoming more 
pronounced for larger spins. Near extremality, the shadow develops the characteristic D-shaped profile, as illustrated 
by the KN shadow in the lower-right panel of Fig.~\ref{fig:Shadows}.

For rEMVBHs, however, we find several notable differences. First, as shown in the upper-left panel of Fig.~\ref{fig:Shadows}, 
at low spin, $\chi=0.1$, the evolution of the EMV shadow closely resembles that of the KN case: the shadow 
remains nearly circular and shrinks as $q$ increases. However, near the extremal KN configuration $(q=0.9947,\chi=0.1)$, 
a mild deformation of the EMV shadow already becomes visible, as shown in the lower-left panel. Specifically, the equatorial 
diameter decreases while the polar diameter increases. This deformation becomes significantly more pronounced at $q=1.2527$, 
where the EMV shadow develops an approximately elliptical shape. As shown in the upper-right panel, increasing $\chi$ 
produces a trend similar to that of KNBHs: the shadow is displaced in the direction of rotation, and the 
displacement increases with $\chi$. But near the extremal KN configuration $(q=0.6887,\chi=0.725)$, the rEMVBHs' 
shadow does not exhibit the pronounced D-shaped deformation seen in the KN case, as illustrated in the lower-right panel. 
We have also compared rEMVBH and KN shadows for several other representative parameter choices. In most cases, 
the two shadows are nearly indistinguishable in shape, differing only slightly in their displacement and overall size.

\begin{figure}[!htbp]
    \centering
    \includegraphics[width=2.975in]{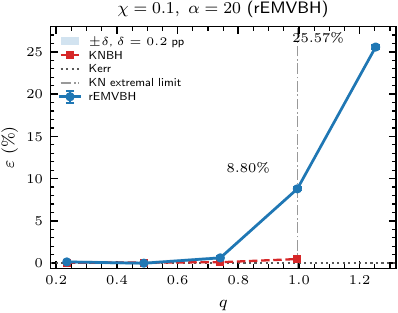}
    \caption{Shadow ellipticity $\varepsilon=H/W-1$ as a function of $q$
    at fixed $\chi=0.1$ for an equatorial observer. Blue circles denote
    rEMVBHs with $\alpha=20$, and red squares denote the corresponding
    analytic KN shadows. The horizontal dotted line gives the Kerr value
    at the same spin. Lines connecting the markers are guides to the eye.
    The vertical dash-dotted line marks the KN extremal bound,
    $q=\sqrt{1-\chi^2}$, beyond which no KN black hole exists at this spin.
        The ellipticity is expressed as a percentage.
    Error bars and the shaded band show $\pm\delta$ with
    $\delta=0.2$ pp, a conservative bound on the combined pipeline and
    resolution uncertainty (measured range $0.15$--$0.25$ pp). At the
    lowest charges shown the ellipticity is consistent with zero within
    this uncertainty, whereas the values at $q=0.9947$ and $q=1.2527$
    lie far outside it.}
    \label{fig:ShadowEllipticityQ}
\end{figure}

To turn these observations into a quantitative criterion, we characterize the deformation by
$\varepsilon\equiv H/W-1$, where $W$ and $H$ are the equatorial and polar diameters of the shadow. Two regimes can be
separated. At fixed $q=0.6887$, where the deformation is driven by rotation alone, the three points
$\chi=0.3$, $0.5$, $0.725$ follow the single power law
\begin{equation}
  \varepsilon = 0.182\,\chi^{1.954},
  \label{eq:eps}
\end{equation}
whose coefficient is about three times that of a Kerr black hole at the same spin, for which the same fit gives
$0.076\,\chi^{2.242}$. An exponent close to two is itself generic for a rotational deformation; what is specific to the
EMV model is the size of the coefficient and its growth as $q\to1$. The second regime is the one that produces a
qualitative signature. As illustrated in Fig.~\ref{fig:ShadowEllipticityQ}, at $\chi=0.10$ the total ellipticity of the vectorized solution at $q=0.9947$ is $8.80\%$,
while the same charge produces only $0.44\%$ in KN (the difference between the KN and Kerr values) and rotation alone
$0.06\%$; the hair therefore adds $8.3\%$, a factor of $19$ beyond what charging a Kerr black hole can achieve, and
the deformation reaches $25.6\%$ at $q=1.2527$, where no KN counterpart exists. A model that merely adds charge to Kerr
has, at this spin, an upper limit of about $0.5\%$; producing $8.8\%$ requires an additional gravitational degree of
freedom. 

Conversely, at the lowest charges we probe the rEMVBH and KN shadows differ by an amount comparable to the
systematics of our ray-tracing pipeline; there the two are therefore
\emph{indistinguishable at our resolution}, and we make no claim of an observable difference.
The comparison is quantitative. The two contributions to our uncertainty --- discretisation, estimated from four
configurations rendered at both $1024^2$ and $2048^2$ ($\max=0.151$ percentage points), and the ray-tracing pipeline
itself, estimated by running native Kerr--Newman through the same pipeline and comparing with the analytic Bardeen
screen ($\max=0.102$ percentage points) --- combine to a resolution in the range $\delta\simeq0.15$--$0.25$
percentage points; we adopt $\delta=0.2$ percentage points as a conservative bound. At $\chi=0.10$ the ellipticity
difference $|\Delta\varepsilon|$ then stays at or below $0.102$ percentage points for $q\le0.6887$
(indistinguishable), and reaches $0.516$ percentage points at $q=0.7407$, $8.30$ at $q=0.9947$ and $25.6$ at
$q=1.2527$ (resolved). At $\chi=0.30$ the difference ranges from $0.74$ to $4.12$ percentage points over the
solutions we probe, so that entire line is resolved. For completeness we 
note that the thermodynamic properties of the vectorized solutions differ more strongly: at high spin the reduced
Hawking temperature differs from that of a Kerr black hole of the same mass by up to $45\%$. This is a thermodynamic
rather than an observational diagnostic, since the Hawking temperature of an astrophysical supermassive black hole lies
far below the cosmic microwave background temperature.

\begin{table}[htbp]
  \centering
    \caption{Physical quantities of the high-charge solution computed with a $100\times50$ grid.
           $\Delta_{\rm Smarr}$ is defined as in Table~\ref{tab:3sols} and is dimensionless.}
  \label{tab:sol4}
  \setlength{\tabcolsep}{2.5pt}
  \begin{tabular}{cccccccccc}
    \hline\hline
        Solution & $q$ & $\chi$ & $T_H$ & $A_H$ & $P$ & $M$ & $J$ & $\Omega_H$ & $\Delta_{\rm Smarr}$ \\
    \hline
        IV & 1.2527 & 0.100 & 0.0091324 & 438.00 & 0.23997 & 5.7808 & 3.3418 & 0.014055 & $2.1092\times10^{-10}$ \\
    \hline\hline
  \end{tabular}
\end{table}

In summary, the most distinctive shadow feature of rEMVBHs appears in the high-charge regime, where the shadow develops 
an approximately elliptical deformation. The corresponding physical quantities for the solution with $q=1.2527$ and 
$\chi=0.1$ are listed in Table~\ref{tab:sol4}, while the profiles of the functions $\mathcal{F}$ are shown in Fig.~\ref{fig:sol4}.

\section{Conclusion}
\label{sec:Conclusion}

In this work, we have numerically constructed rotating black hole solutions in the EMV model by adopting suitable ansatz and 
boundary conditions. We have investigated the contributions of the vector hair to the Komar mass, the domain of existence 
of these solutions, and their shadow properties.

Using the Iyer--Wald covariant phase space formalism, we have derived the Smarr relation for rEMVBHs. Combining this result 
with the Komar integral identities yields $M_V=2\Omega_HJ_V$, relating the Komar mass contribution of the vector hair to 
its angular momentum. The fractional contribution of the vector hair to the total mass increases with $\chi$ and 
decreases as the coupling constant $\alpha$ becomes larger.
We have further explored the domain of existence of rEMVBHs in the $q$--$\chi$ parameter space at $\alpha=20$. 
These solutions can extend beyond the KN extremality bound, $q^2+\chi^2=1$, and even reach the region $q>1$. 
Finally, our calculations of black hole shadows reveal that, although rEMVBH shadows closely resemble their 
KN counterparts for many of the configurations examined, distinctive deformations can arise at large charge. 
In particular, the solution with $q=1.2527$ and $\chi=0.1$ exhibits an approximately elliptical shadow for an equatorial observer.
The deformation obeys the single scaling law of Eq.~\eqref{eq:eps}, and at the lowest charges we probe the
vectorized and Kerr--Newman shadows are indistinguishable at our resolution; observational statements about the
model should be restricted to the overcharged regime where the deviation reaches $8.8\%$ and beyond.

The question posed in the Introduction was whether the undercharging of spherical vectorized black holes is a property of
the vector field or of spherical symmetry, and whether rotation can therefore lift the three degeneracies that spherical
symmetry imposes --- the vanishing Komar mass of the hair, the identity $M_V=2\Omega_HJ_V$ reducing to $0=0$, and the
ceiling on the charge. The results above answer it in the affirmative, and the answer has a rigid structure that is worth
stating in physical terms. Rotation makes the hair weigh, but only through the rotational channel: $M_V$ is not an
independent parameter of the solution but a second-order invariant of the rotational response, exactly locked by
$M_V=2\Omega_HJ_V$, so that the hair owns no extensive quantity of its own and no independent hair term appears in the
Smarr relation. The same coupling that makes the hair weigh also displaces the charge ceiling, and it does so only above a
threshold in the coupling. The resulting deformation is geometric, and it is a signature of the hair rather than of the
charge: no solution that merely charges a Kerr black hole can reproduce the ellipticity we find.

Beyond the shadow boundary, suitable emission models could be adopted to investigate the shapes, widths, and relative 
fluxes of photon subrings, exploring whether these finer structures provide signatures of vector hair beyond those 
encoded in the shadow contour~\cite{Johnson:2019ljv,Staelens:2023jgr}. In addition, a caveat is in order 
concerning stability~\cite{Berti:2009kk,Khoo:2024agm,Xiong:2024urw}. Our analysis is restricted to stationary, axisymmetric and asymptotically
flat configurations. The existence of a stationary solution and the stability of its fluctuations are logically
independent properties: the former is a property of the elliptic boundary-value problem solved here, whereas the
latter is a property of the hyperbolic sector, i.e.\ of the quadratic action for perturbations about that solution.
Recent work has shown that the tachyonic instability responsible for spontaneous vectorization is generically
accompanied by ghost- or gradient-type instabilities in the linearized theory~\cite{Pizzuti:2023eyt,Clough:2022ghost,Chiang:2026nogo},
and that non-linear evolution does not remove them~\cite{Clough:2022ghost}. We do not attempt to resolve this question
here, and we make no claim about the dynamical formation of the black holes constructed above.

A natural continuation is therefore the stability analysis of the rotating branch. Expanding the EMV action to
quadratic order about a hairy background $\bar{\mathcal{B}}_\mu$ gives
\begin{equation}
  S_2=\int d^4x\,\sqrt{-g}\left[-\tfrac14\,\delta V_{ab}\delta V^{ab}-\tfrac12 M^{ab}\,\delta \mathcal{B}_a\delta \mathcal{B}_b\right],
  \qquad
  M^{ab}=m_{\rm eff}^2\left(g^{ab}+2\alpha\,\bar{\mathcal{B}}^a\bar{\mathcal{B}}^b\right),
\end{equation}
with $m_{\rm eff}^2\equiv\tfrac12\alpha\,e^{\alpha|\bar{\mathcal{B}}|^2}F_{ab}F^{ab}$, so that the effective mass is negative,
and vectorization therefore possible, wherever the electromagnetic invariant is electric dominated.

For the rotating backgrounds constructed here the premises of the criteria of Ref.~\cite{Chiang:2026nogo} fail, and
this can be shown constructively. Those criteria rest on a purely quadratic dispersion relation, which requires the
absence of off-diagonal terms in the time-like direction. Writing the dispersion matrix of vector perturbations as
$\mathcal D^{bd}=k^2g^{bd}-k^bk^d+M^{bd}$ with $k_a=(-\omega,k,0,\ell)$, its determinant factorizes as
\begin{equation}
  \det\mathcal D=\det\!\left(k^2G+M\right)\left[1-k_a\left(k^2G+M\right)^{-1}_{ab}k_b\right],
\end{equation}
so that $\mathcal D=\omega^2A_2+\omega A_1+A_0$ with $A_1\neq0$ whenever $\ell\neq0$ (or $k\neq0$) together with
$g^{t\varphi}\neq0$ or $M^{ti}\neq0$; the determinant then contains odd powers of $\omega$, in contrast to the purely
quadratic case. This is not a marginal effect: for $88.6\%$ of the $47\,512$ archived $\alpha=20$ solutions the largest
off-diagonal component of $M^{ab}$ reaches at least $10\%$ of its diagonal scale, a further $10.8\%$ lie in a marginal
range, and only $0.6\%$ satisfy the premise.

It is important to state what this does and does not mean. A negative effective mass is not by itself a pathology of
the model: it is the very mechanism that generates the hair, and it is present for the spherically symmetric
vectorized solutions as well. What we find is that the \emph{tachyon-free} region, in the sense of
Ref.~\cite{Chiang:2026nogo}, is empty on these backgrounds. Solving the complete dispersion relation for three
representative solutions with off-diagonal strength $V_{\rm od,rel}=0.088$, $0.444$ and $0.768$, we obtain purely
imaginary $\omega$ at zero momentum in $14$ of the $15$ sampled points, so the tachyonic mode is present; increasing
the off-diagonal components by a factor of $8.7$ does not open a tachyon-free window, and they are therefore not a
rescue mechanism. A sign flip of the eigenvalue of $M^{ab}$ along $\bar{\mathcal{B}}$ requires $2\alpha|b^2|>1$, where
$b^2\equiv\bar{\mathcal{B}}_a\bar{\mathcal{B}}^a$, and is absent for $\alpha\gtrsim15$; it does not operate in the $\alpha=20$ sample
analysed here. Two questions remain open. First, gradient stability at large $k$ is not settled: within the sampled
range ${\rm Im}\,\omega$ does not grow monotonically with $k$, but the large-$k$ limit requires a dedicated treatment
because the background varies by many orders of magnitude. Second, and more importantly, ghost freedom cannot be
decided from the dispersion relation alone, which is invariant under $\mathcal D\to\Lambda\mathcal D$ whereas the sign
of the kinetic term is not; a Hamiltonian or kinetic-matrix analysis would be required. We therefore do not claim
either to confirm or to refute the theorem of Ref.~\cite{Chiang:2026nogo}: on our backgrounds its premises are not
satisfied, and the dynamical viability of the rotating branch remains an open question.

What this establishes is a statement about the stationary solution space, and it is structural:
in the EMV model the Kerr--Newman bound is a property of one sector of the solution space, not of
the field equations themselves. At $\alpha=20$ the spherical existence line terminates at
$q_{\max}^{\rm sph}=0.6947$; the rotating families constructed here grow out of that line, to within
$0.005\%$, and cross $q^2+\chi^2=1$, where $q\equiv Q/M$ and $\chi\equiv J/M^2$ are the same ADM
ratios that enter the Kerr--Newman bound. These are black holes, not naked singularities: every
archived solution has $T_H>0$ and $A_H>0$, and the solution of Table~\ref{tab:sol4} has a horizon of
finite area and nonzero temperature at $q^2+\chi^2=1.58$, with a dimensionless Smarr residual of
$2\times10^{-10}$. Whether such objects are populated in the Universe is a separate, observational
question. The discharge argument of Ref.~\cite{Coviello:2025viab} --- for an extremal hole the
lifetime becomes untenable below $\sim10^{14}M_\odot$ --- constrains their abundance, not the
structure of the solution space, and the two are logically independent: even if this branch were
stable, it would still be rare. Two features specific to the present model are worth recording in
this context. The vector charge $P$ is a hidden charge, which is precisely the case in which
Ref.~\cite{Coviello:2025viab} allows the standard discharge bounds to be evaded; and because the
hair is sourced by $F^2$, discharging $Q$ would remove the hair altogether, leaving a nearly bald
Kerr hole rather than a neighbouring EMV solution. We claim neither abundance nor detection: the
shadow deformation of Sec.~\ref{sec:Shadows} is a prediction of what an observer would record if
such a black hole were present.

\begin{acknowledgments}\sloppy
P.\,L. thanks Y.\,H. Cha, Y.\,E. Liu and B. Liu for their help during this work.
This work is supported by the Natural Science Foundation of China under Grant
No. 12475054, and by the Guangdong Basic and Applied Basic Research
Foundation under Grant No. 2025A1515012063.
\end{acknowledgments}

\noindent\textit{Data and code availability.}---The public repository
\href{https://github.com/physicsuniverse/rEMVBH-reproduction}
{\texttt{github.com/physicsuniverse/\allowbreak rEMVBH-reproduction}}~\cite{Guo:2026repro} contains the
derived equations, the archived rotating solutions, the per-point residual diagnostics used to
define the accuracy-qualified region, the unit tests on known limits, and the plotting code that
regenerates every figure in this paper from the stored data.

\appendix
\section{Smarr Relation of the EMV Model}
\label{app:Smarr}

Wald and Iyer developed the covariant phase-space formalism for general diffeomorphism-invariant 
theories of gravity. By writing the variation of the Lagrangian as
\begin{equation}
  \delta\mathbf{L}=\mathbf{E}_{\phi}\delta\phi+d\mathbf{\Theta}\left(\phi,\delta\phi\right),
\end{equation}
they constructed the Noether current $\mathbf{J}[\xi]$ and the associated Noether 
charge $\mathbf{Q}[\xi]$, and showed that the entropy of a stationary black hole 
with a bifurcate Killing horizon can be expressed as an integral of the Noether 
charge over the horizon \cite{Wald:1993nt,Iyer:1994ys}.

Based on this formalism, Ref.~\cite{Liberati:2015xcp} proposed a method for deriving 
the Smarr relation directly from the Lagrangian. For a stationary black hole 
generated by a Killing vector field $\xi$, the corresponding Noether current is
\begin{equation}
  \mathbf{J}[\xi] = \mathbf{\Theta}\left(\phi,\pounds_{\xi}\phi\right) - \iota_\xi\mathbf{L}.
\end{equation}
With $\boldsymbol{\epsilon}$ the four-dimensional volume form and $\iota$ the interior product,
$\iota_\xi\mathbf{L}=\mathcal{L}\,\iota_\xi\boldsymbol{\epsilon}$ is a three-form. 
On shell, the Noether current is closed and can therefore be locally expressed as 
the exterior derivative of the Noether charge $\mathbf{Q}[\xi]$. Here, “on shell” 
means that the background fields $\phi$ satisfy the equations of motion, $\mathbf{E}_{\phi}=0$. 
Moreover, since $\xi$ generates a symmetry of the background fields, 
$\pounds_{\xi}\phi=0$, the symplectic-potential term $\mathbf{\Theta}\left(\phi,\pounds_{\xi}\phi\right)$ vanishes. 
Consequently,
\begin{equation}
  d\mathbf{Q}[\xi] \doteq \mathbf{J}[\xi] \doteq -\iota_\xi\mathbf{L},
\end{equation}
where $\doteq$ denotes equality on shell. Integrating this identity over a spacelike 
hypersurface extending from the black hole horizon to spatial infinity and applying 
Stokes' theorem converts the left-hand side into boundary integrals of the Noether charge.
\begin{equation}
  \label{eqofQtoSmarr}
  \int_{\partial\Sigma} \mathbf{Q}[\xi] +\int_\Sigma \iota_\xi\mathbf{L} \doteq 0.
\end{equation}
The Smarr relation is then obtained by identifying the horizon contribution, 
the conserved charge at infinity, and the bulk contribution from the on-shell Lagrangian.

We now apply this method to the EMV model. Since Eq.~\eqref{eqofQtoSmarr} is an on-shell identity, 
the equations of motion can be used to simplify the Noether charge and the second volume integral. 
We directly present the Noether current:
\begin{equation}
  \label{NoetherJ}
  \mathbf{J}_{bcd} \doteq \frac{1}{8\pi} \epsilon_{abcd} \nabla_e 
                   \left( \frac{1}{2} \left(\nabla^{e}\xi^{a} - \nabla^{a}\xi^{e}\right) + 2 U F^{ea} \mathcal{A}_f\xi^f + 2 V^{ea} \mathcal{B}_f\xi^f \right).
\end{equation}
We need to identify the differential form whose exterior derivative gives the expression above. 
In this regard, it is useful to apply the codifferential to this expression \cite{Bartelmann_2019,Peng:2019woh}. 
We denote the expression in braces on the right-hand side of Eq.~\eqref{NoetherJ} by the 
antisymmetric tensor $\mathbf{X}^{ea}$. It then follows that the Noether current is given by 
$\mathbf{J}\doteq-\star\delta\mathbf{X}$, where $\star$ denotes the Hodge dual 
and $\delta$ represents the codifferential. 
Therefore, the corresponding Noether charge is
\begin{equation}
  \label{NoetherQ}
  \mathbf{Q}_{cd} = -\frac{1}{16\pi} \epsilon_{abcd} 
                    \left( \nabla^a\xi^b + 2 U F^{ab} \mathcal{A}_e\xi^e + 2 V^{ab} \mathcal{B}_e\xi^e \right).
\end{equation}
Integrating Eq.~\eqref{NoetherQ} over $\partial\Sigma=S_\infty-H$, we obtain
\begin{equation}
  \label{intofQ}
  \int_{\partial\Sigma} \mathbf{Q}[\xi] = \frac{1}{2} M - \Omega_H J - Q \Phi - T_H S - \frac{1}{8\pi} \int_{\partial\Sigma} \epsilon_{abcd} V^{ab} \mathcal{B}_e\xi^e.
\end{equation}
Under the boundary conditions adopted in this work, $\mathcal{B}_e\xi^e$ vanishes both on the event horizon and at 
spatial infinity. Consequently, the last integral term in Eq.~\eqref{intofQ} vanishes. As for the second volume integral in 
Eq.~\eqref{eqofQtoSmarr}, the equations of motion allow us to rewrite Lagrangian $\mathcal{L}$ as:
\begin{equation}
\begin{aligned}
  \mathcal{L} 
  &\doteq \frac{1}{16\pi} \left(-2 T - U F^{ab} F_{ab} - V^{ab} V_{ab}\right) \\
  &= - \frac{1}{8\pi} \left( \frac{1}{2} U' F^2 \mathcal{B}^b \mathcal{B}_b + U F^{ab} \nabla_a \mathcal{A}_b + V^{ab} \nabla_a \mathcal{B}_b\right) \\
  &\doteq -\frac{1}{8\pi} \left( \nabla_a\left(\mathcal{B}_b V^{ab}\right) + \nabla_a \left(U \mathcal{A}_b F^{ab}\right) \right)
\end{aligned}
\end{equation}
The first $\doteq$ follows from Eq.~\eqref{eommetric}, where the Ricci scalar is replaced by the trace of 
the energy--momentum tensor. The second $\doteq$ is obtained by simultaneously using Eqs.~\eqref{eomF} and \eqref{eomV}. 
Consequently, $\mathcal{L}$ can be rewritten as $\nabla_a K^a$.

The passage from this on-shell identity to a boundary term is most transparent in the
language of differential forms, and needs no separate divergence theorem. Pulling it
back to the hypersurface $\Sigma$ of constant $t$ leaves only the $t^a$ contribution: since
$\xi^a=t^a+\Omega_H\varphi^a$, the contraction with $\varphi^a$ carries a factor of $dt$
whose pullback to $\Sigma$ vanishes. Using $\pounds_\xi\boldsymbol{\epsilon}=0$ (the Killing
condition) and the Cartan identity
$\pounds_\xi\iota_K\boldsymbol{\epsilon}
=\iota_{[\xi,K]}\boldsymbol{\epsilon}+\iota_K\pounds_\xi\boldsymbol{\epsilon}$,
together with $[\xi,K]=0$ --- which holds because $K^a$ is built covariantly out of the
fields $g$, $\mathcal{A}$ and $\mathcal{B}$ that are themselves invariant under
$\xi$ --- one obtains
\begin{equation}
\label{contraction}
  \left(\nabla_aK^a\right)\iota_\xi\boldsymbol{\epsilon}
  = -\,d\!\left(\iota_\xi\iota_K\boldsymbol{\epsilon}\right),
\end{equation}
in which both sides are three-forms and the right-hand side is a pure total derivative.
Note that Eq.~\eqref{contraction} involves no $d$(four-form): such a term would be a
five-form and would vanish identically in four dimensions. The boundary integrand is
therefore the \emph{product} $\iota_\xi\iota_K\boldsymbol{\epsilon}$, whose
$(\theta,\varphi)$ component is $-K^r\sqrt{-g}$, so that
\begin{equation}
\label{intofL}
  \int_\Sigma \iota_\xi\mathbf{L}
  =  \int_{S_\infty} \sqrt{-g}\,K^r\,d\theta d\varphi
  - \int_{\mathcal{H}} \sqrt{-g}\,K^r\,d\theta d\varphi ,
\end{equation}
where stationarity and axisymmetry eliminate the temporal and azimuthal derivatives, while
regularity on the symmetry axis eliminates the polar boundary terms. The remaining radial
boundary integral is understood as the contribution at spatial infinity minus that at the
horizon.

To evaluate this integral, we substitute the metric and
field ansatz into
$K^r=-(8\pi)^{-1}
(\mathcal{B}_bV^{rb}+U\mathcal{A}_bF^{rb})$
and express it in terms of the numerical functions
$\mathcal{F}=\{B_t,B_\varphi,A_t,A_\varphi,f,g,h,W\}$
and their derivatives. Applying the boundary conditions,
together with horizon regularity, shows that the
$\mathcal{B}$ contribution to the boundary integral vanishes
both at the horizon and at spatial infinity. The
$\mathcal{A}$ contribution also vanishes at the horizon
in the gauge adopted here. At spatial infinity,
$U\to1$, $\mathcal{A}_t=\Phi-Q/r+O(r^{-2})$,
$F^{rt}=-Q/r^2+O(r^{-3})$, and
$\sqrt{-g}=r^2\sin\theta[1+O(r^{-1})]$.
Consequently,
$\sqrt{-g}K^r\to Q\Phi\sin\theta/(8\pi)$,
whose integral over the sphere yields $Q\Phi/2$.
Substituting this result, together with
Eq.~\eqref{intofQ}, into Eq.~\eqref{eqofQtoSmarr},
we obtain the Smarr relation Eq.~\eqref{SmarrnoV}.

\section{Profiles of the Numerical Solutions}
\label{app:solplots}

\begin{figure}[!htbp]
    \centering
    \includegraphics[width=5.95in]{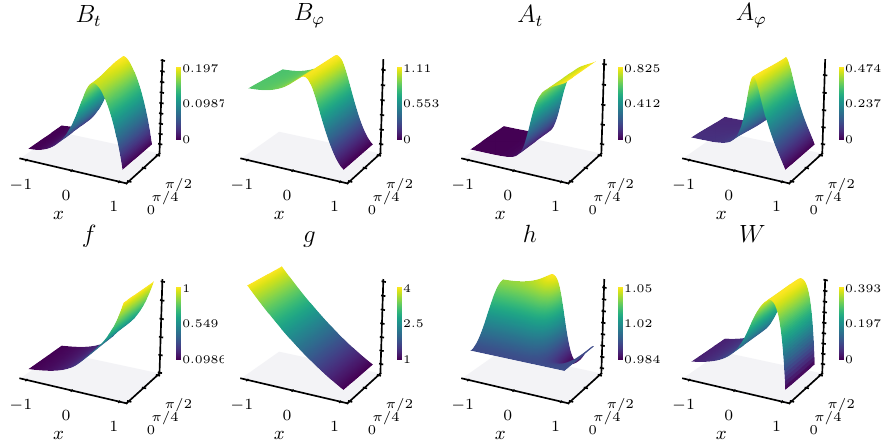}
    \caption{Profiles of the functions $\mathcal{F}$ for numerical solution I listed in Table~\ref{tab:3sols}. The solution is reflection-symmetric about the equatorial plane, $\theta\to\pi-\theta$; $\theta\in[0,\pi/2]$ is the fundamental domain we solve, and is what we display.}
    \label{fig:sol1}
\end{figure}

\begin{figure}[!htbp]
    \centering
    \includegraphics[width=5.95in]{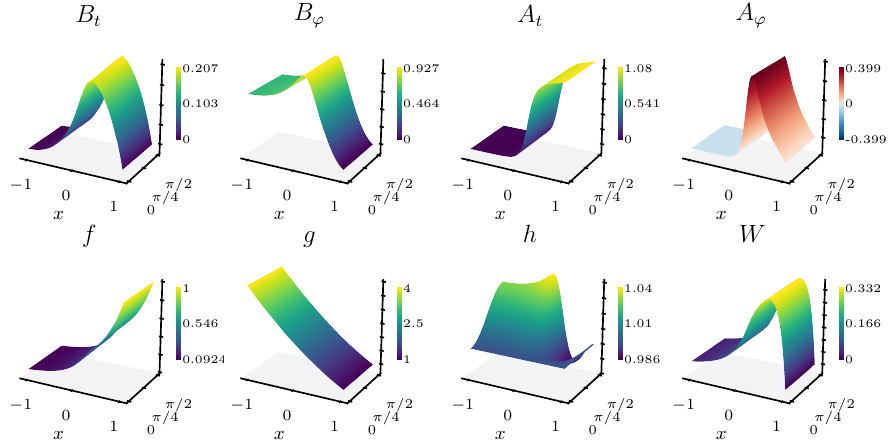}
    \caption{Profiles of the functions $\mathcal{F}$ for numerical solution II listed in Table~\ref{tab:3sols}. The solution is reflection-symmetric about the equatorial plane, $\theta\to\pi-\theta$; $\theta\in[0,\pi/2]$ is the fundamental domain we solve, and is what we display.}
    \label{fig:sol2}
\end{figure}

\begin{figure}[!htbp]
    \centering
    \includegraphics[width=5.95in]{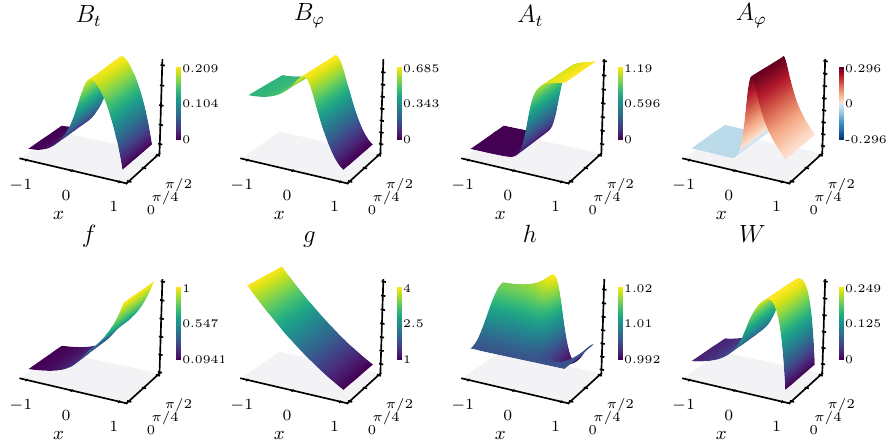}
    \caption{Profiles of the functions $\mathcal{F}$ for numerical solution III listed in Table~\ref{tab:3sols}. The solution is reflection-symmetric about the equatorial plane, $\theta\to\pi-\theta$; $\theta\in[0,\pi/2]$ is the fundamental domain we solve, and is what we display.}
    \label{fig:sol3}
\end{figure}

\begin{figure}[!htbp]
    \centering
    \includegraphics[width=5.95in]{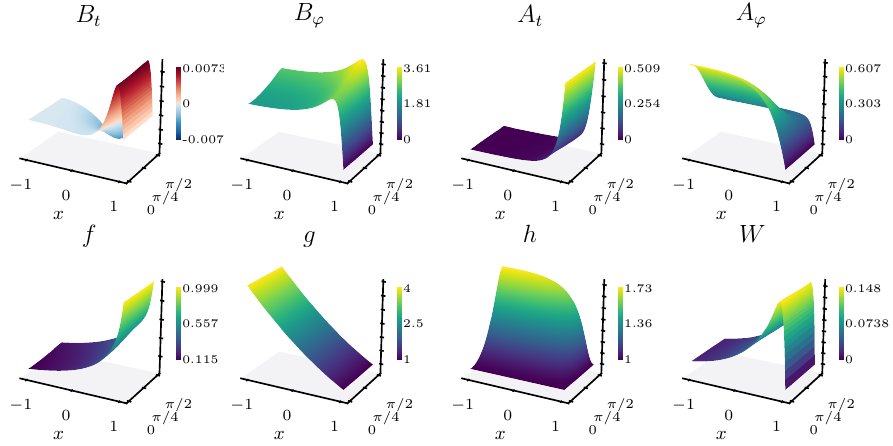}
    \caption{Profiles of the functions $\mathcal{F}$ for numerical solution IV listed in Table~\ref{tab:sol4}. The solution is reflection-symmetric about the equatorial plane, $\theta\to\pi-\theta$; $\theta\in[0,\pi/2]$ is the fundamental domain we solve, and is what we display.}
    \label{fig:sol4}
\end{figure}

\begin{figure}[!htbp]
    \centering
    \includegraphics[width=5.95in]{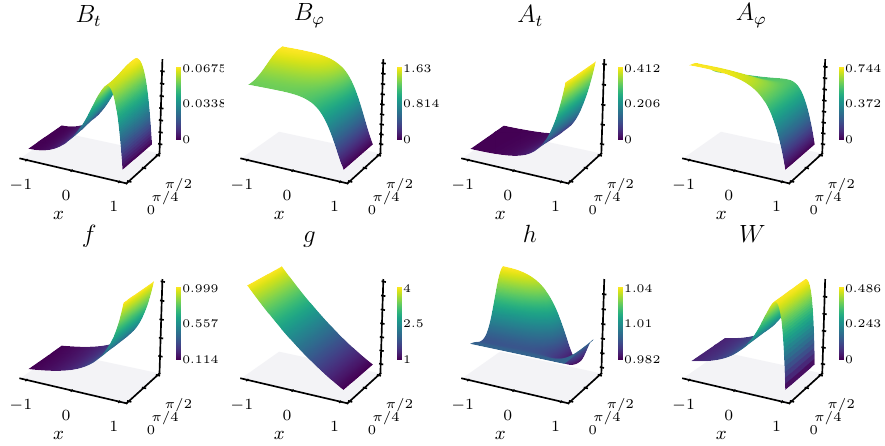}
    \caption{Profiles of the eight functions
             $\{B_t,B_\varphi,A_t,A_\varphi,f,g,h,W\}$ for the numerical solution with
             $q=0.6937$, $\chi=0.500$ and $\alpha=20$, computed on a $60\times20$ grid.
             The surfaces are drawn on the numerical coordinates actually used by the
             solver, $x=1-2r_H/r\in[-1,1]$, and have been spectrally interpolated onto a
             finer grid for rendering. The solution is reflection-symmetric about the equatorial
             plane, $\theta\to\pi-\theta$; $\theta\in[0,\pi/2]$ is the fundamental domain we solve,
             and is what we display. This is not an extrapolation, since the spectral basis
             $\{\cos 2j\theta\}$ is even under $\theta\to\pi-\theta$.}
    \label{fig:sol_profiles}
\end{figure}

\begin{figure}[!htbp]
    \centering
    \includegraphics[width=5.95in]{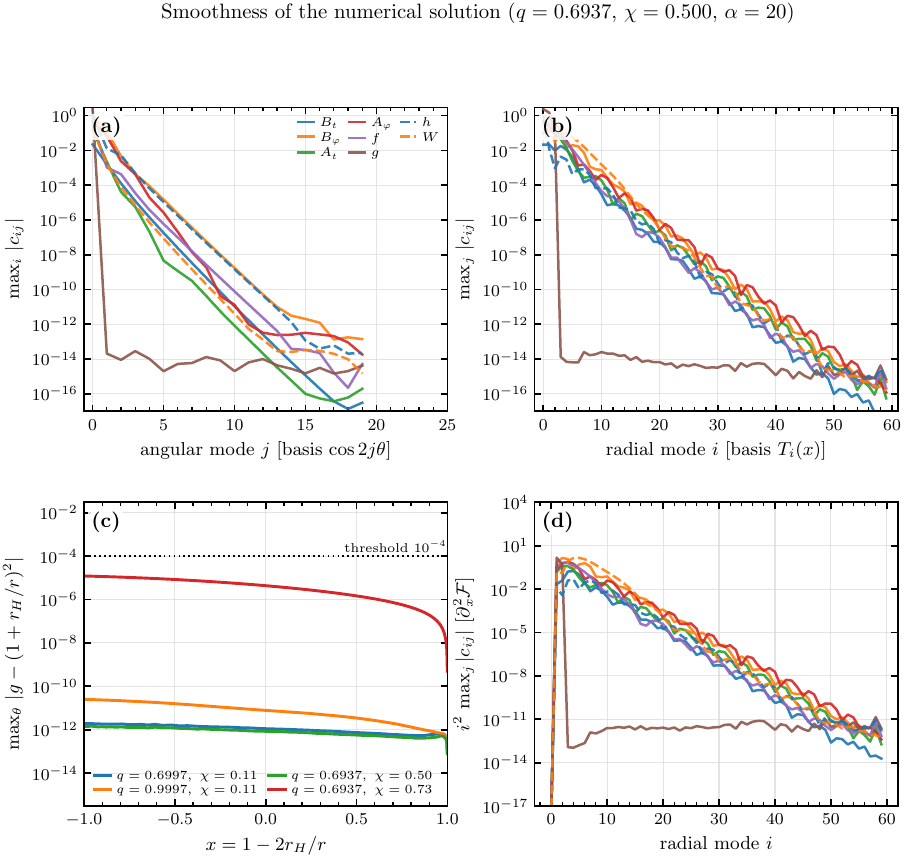}
    \caption{Smoothness of the numerical solution with $q=0.6937$ and $\chi=0.500$.
             Top: magnitude of the angular \textbf{(a)} and radial \textbf{(b)} spectral
             coefficients $c_{ij}$ in the expansion
             $\mathcal{F}=\sum_{i,j}c_{ij}T_i(x)\cos(2j\theta)$, showing geometric decay
             down to machine precision. Bottom: \textbf{(c)} deviation of the computed
             $g$ from the analytic expression $g=(1+r_H/r)^2$ for several solutions;
             \textbf{(d)} the derivative spectrum $i^2\max_j|c_{ij}|$, showing that second
             derivatives are equally well resolved. Note that the angular coefficients of
             $g$ vanish for $j\geq1$ at the $10^{-14}$ level, which numerically confirms
             the analytic result $g=(1+r_H/r)^2$.}
    \label{fig:smoothness}
\end{figure}

We collect here the profiles of the eight functions $\mathcal{F}=\{B_t,B_\varphi,A_t,A_\varphi,f,g,h,W\}$ for the
representative solutions discussed in the main text. Figures~\ref{fig:sol1}, \ref{fig:sol2} and \ref{fig:sol3} show
the three $100\times50$ solutions listed in Table~\ref{tab:3sols}, and Fig.~\ref{fig:sol4} the high-charge solution of
Table~\ref{tab:sol4}. Figure~\ref{fig:sol_profiles} gives the same set of functions for the $60\times20$ solution with
$q=0.6937$, $\chi=0.500$, and Fig.~\ref{fig:smoothness} quantifies the accuracy of the spectral representation for
that solution.

\clearpage

\bibliographystyle{apsrev4-2}
\bibliography{rotatingEMVcite}

\end{document}